\documentclass[pdftex]{aa}  
\usepackage{hhline}
\usepackage{hyperref}
\usepackage[utf8]{inputenc}
\usepackage{amssymb}
\usepackage{lscape}
\usepackage{graphicx}
\usepackage{float}
\usepackage{natbib}
\usepackage{txfonts}

\begin{document}

   \title{Polarimetry of M-type asteroids in the context of their surface composition}
   \titlerunning{Polarimetry of M-type asteroids}
     \author
          {I. Belskaya,
          \inst{1}
          A. Berdyugin,
          \inst{2}
          Yu. Krugly,
          \inst{1}
          Z. Donchev,
          \inst{3}
           A. Sergeyev,
          \inst{1}
          R. Gil-Hutton,
          \inst{4}
          S. Mykhailova,
          \inst{1}
          T. Bonev,
          \inst{3}
          V.~Piirola,
          \inst{2}
          S. Berdyugina,
          \inst{5}
           M. Kagitani,
            \inst{6}
           T. Sakanoi\inst{6}
          }
\authorrunning{I. Belskaya et al.}

   \institute{V. N. Karazin Kharkiv National University, 4 Svobody Sq., Kharkiv, 61022, Ukraine
         \and
             Department of Physics and Astronomy, FI-20014 University of Turku, Finland 
           \and
              Institute of Astronomy and NAO, Bulgarian Academy of Sciences, Sofia, Bulgaria
          \and
              Planetary Science Group, Universidad Nacional de San Juan and CONICET, Av. José I. de la Roza 590 (O), J5402DCS Rivadavia, San Juan, Argentina 
           \and
               Leibniz-Insititut für Sonnenphysic, D-79104, Freiburg, Germany
          \and
               Graduate School of Science, Tohoku University, Aoba-ku, Sendai 980-8578, Japan      
              }

   \date{Received November 30, 2021; accepted }

  \abstract
  % context heading (optional)
  % {} leave it empty if necessary  
   {\textit{Aims.} We aim to investigate how polarimetric observations can improve our understanding of the nature and diversity of M/X-type asteroids.
   
   \textit{Methods.} Polarimetric observations of the selected M/X-type asteroids were carried out at the Tohoku 0.6-m telescope at Haleakala Observatory, Hawaii (simultaneously in BVR filters), the 2-m telescope of the Bulgarian National Astronomical Observatory in Rozhen (in R filter), and the 2.15-m telescope of the Complejo Astronómico El Leoncito (CASLEO), Argentina (in V filter). We analysed the polarimetric characteristics of M/X-type asteroids along with the available data obtained by other techniques.
   
   \textit{Results.} New polarimetric observations of 22 M/X-type asteroids  combined with published observations provide a data set of 41 asteroids for which the depth of a negative polarisation branch and/or inversion angle were determined. We found that the depth of the negative polarisation branch tends to increase with decreasing steepness of the near-infrared spectra. Asteroids with a deeper negative polarisation branch tend to have a higher radar circular polarisation ratio. We show that, based on the relationship of the depth of the negative polarisation branch and inversion angle, two main sub-types can be distinguished among M-type asteroids. We suggest that these groups may be related to different surface compositions similar to (1) irons and stony-irons and (2) enstatite and iron-rich carbonaceous chondrites.
    }

   \keywords{minor planets, asteroids: general -- techniques: polarimetric
               }

   \maketitle
%
%-------------------------------------------------------------------

\section{Introduction}

   Using the letter M (metal) to classify several asteroids was proposed by  \citet{zellner} due to the similarity of their polarimetric and spectral properties to iron meteorites. However, possible spectral meteorite analogues of these asteroids included not only iron meteorites but also some types of enstatite chondrites (\citealt{chapman}). 'M type' was one of seven major types in Tholen’s taxonomy that was well separated from other types by featureless spectra and moderate surface albedo (\citealt{tholen}). In recent classifications, M type is a part of the X complex, which includes all asteroids with featureless spectra regardless of their albedo (\citealt{bus}, \citealt{lazzaro}, \citealt{demeo}).
    
    Initially, it was believed that M-type asteroids could be the remnants of the metal cores of differentiated planetesimals, and this caused a great interest in their study. Numerous observations of M-type asteroids by various techniques revealed that M-type includes asteroids of diverse composition. Rivkin et al. (\citeyear{rivkin}; \citeyear{rivkin00}) found absorption features at 3 $\mu$m that were inconsistent with a metal-dominated composition for 10 out of 27 M-type asteroids. Spectroscopic surveys of M/X-type asteroids revealed a variety of spectral behaviours, as well as weak absorption features in the near-infrared wavelength range (\citealt{clark}, \citealt{fornasier}, \citeyear{fornasier11}, \citealt{ockert08}, \citeyear{ockert10}, \citealt{hardersen11}, \citealt{neeley}). In total, the near-infrared spectra were measured for about 45 M-type asteroids. Spectral data obtained for the same asteroid by different authors often showed inconsistencies in the identification of weak absorption bands (\citealt{fornasier}, \citealt{hardersen11}). Radar studies of M/X-type asteroids were summarised by Shepard et al. (\citeyear{shepard10}, \citeyear{shepard15}). They found that the radar albedos for 11 of 29 measured M-type asteroids matched a metal-dominated composition. A joint analysis of spectroscopic and radar observations of M-type asteroids revealed some inconsistency in their compositional interpretations (\citealt{neeley}, \citealt{shepard15}). 
    
   One more remote sensing technique which efficiently constrains physical properties of atmosphereless bodies surfaces is polarimetry.
 The degree of linear polarisation $\textit{P}_{r}$ of sunlight scattered by an asteroid’s surface is usually defined in terms of the differences between the intensities of the components of the light beam polarised along the planes perpendicular and parallel to the scattering plane. At small phase angles, the latter component dominates, which leads to the so-called negative polarisation branch in the phase-angle dependence of linear polarisation $\textit{P}_{r}$ ($\alpha$). The parameter $\textit{P}_{min}$ characterises the depth of the negative polarisation branch, and the inversion angle $\alpha_{inv}$, at which the sign of $\textit{P}_{r}$ changes, characterises its width.  The relationship between $\textit{P}_{min}$ and $\alpha_{inv}$ is considered as indicative of an asteroid’s surface texture and composition (\citealt{doll89}, \citealt{belskaya17}). \citet{doll79} concluded that the polarimetric parameters of M-type asteroids are compatible with metallic bodies. This conclusion was made by comparing the polarimetric parameters of four M-type asteroids with laboratory data for iron meteorites and various metallic samples. \citet{lupish89} showed that such an interpretation was not unique since some types of ordinary chondrites exhibited polarimetric properties similar to M-type asteroids. \citet{gilhutton} carried out the polarimetric survey of M-type asteroids and obtained data on 26 asteroids. Combining obtained and previously published data, the polarimetric parameters $\textit{P}_{min}$ and/or $\alpha_{inv}$ were estimated for only 12 asteroids. \citet{gilhutton} found that several of them have an M-type controversial taxonomic classification, and the group of asteroids showing in their spectra's 3 $\mu$m hydration band  (\citealt{rivkin}, \citeyear{rivkin00}) has polarimetric parameters that are different from those without that feature. 
   
 Recent interest in studying metal-rich asteroids increased after the largest M-type asteroid, (16) Psyche, was selected as a target of the forthcoming NASA space mission, which is expected to launch in 2022. Intense pre-flight observations of (16) Psyche have not resolved contradictions in understanding Psyche’s composition (e.g. \citealt{elkins}).
   
The present work aims to explore how polarimetry can complement previous analysis of M/X-type asteroids based on spectral and radar data. The results of new polarimetric observations are presented in Section 2. Section 3 describes the determination of polarimetric parameters combining new and published data on M-type asteroids. Section 4 is devoted to the results of searching for correlations between polarimetric parameters and other characteristics. Finally, in Section 5 we try to answer the following question: what can polarimetry tell us about the nature of M-type asteroids?  

\section{Observations}

   For our observations, we selected targets from the list of asteroids classified as M types by \citet{tholen} or as X complexes in other available classifications (\citealt{bus}, \citealt{lazzaro}, \citealt{demeo}), which have moderate surface albedo from 0.1 to 0.35 according to Akari data (\citealt{usui}, \citealt{ali}) and/or WISE data (\citealt{masiero},  \citealt{mainzer}). The main purpose of our observations was to determine the values of polarimetric parameters $\textit{P}_{min}$ and the inversion angle $\alpha_{min}$. Observations were carried out using telescopes at three different observational sites:  the remotely controlled Tohoku 0.6-m telescope at the Haleakala Observatory, Hawaii, the 2-m telescope of the Bulgarian National Astronomical Observatory in Rozhen, and the 2.15-m telescope of the CASLEO, Argentina. We obtained polarimetric measurements of 22 asteroids from May 2018 to November 2021. Observations of 11 asteroids were carried out simultaneously in BVR filters, while 11 other asteroids were observed either in the V or R filter. Below, we briefly characterise the specificities of the observations carried out with each instrument.
   
   \textbf{Observations at the Haleakala Observatory} were carried out at the Tohoku 0.6-m telescope equipped with high-precision Dipol-2 polarimeter. This instrument has three CCD cameras and is capable of simultaneously measuring linear polarisation in the BVR pass bands. To control instrumental polarisation, observations of non-polarised (more than 20 stars per typical observing run) and strongly polarised standard stars have been carried out. The magnitude of instrumental polarisation in all pass-bands is $ \leq 10^{-4}$.  The description of the polarimeter, data acquisition, and data reduction can be found in  \citet{piirola} and  \citet{piirola20}.
   
   \textbf{Observations at the Bulgarian National Astronomical Observatory in Rozhen}  were conducted with the two-channel focal reducer (FoReRo2) (\citealt{jockers}) installed on the Cassegrain focus of the 2-m telescope at the Rozhen Observatory. In polarimetric mode, FoReRo2 is equipped with a $\lambda/2$ retarder wave plate. A Wollaston prism is placed in the parallel beam before the color divider, and thus it feeds both channels, the red and the blue, simultaneously. To measure the linear polarisation of asteroids, we followed the technique described by \citet{bagnulo06}. Observations were carried out in the R band of the  Johnson-Cousins system (R-channel) with an Andor’s iKon-L 936 CCD camera. We obtained a series of CCD frames at different half-wave plate positions rotated with 22.5\textdegree~steps. The telescope tracking was used to keep a moving asteroid in the centre of the apertures. Each night, we measured two or more polarimetric standard stars with high and low polarisation to control the instrumental polarisation. The instrumental polarisation was less than 0.02\%, and the position angle deviation was about 0.5\textdegree.\, The observational data were processed using the standard procedure \citep{bagnulo06}.
   
  \textbf{Observations at CASLEO}  were carried out with the 2.15 m telescope and using the CASPOL polarimeter. The CASPOL instrument is a polarisation unit inserted in front of a CCD camera that allows high-precision imaging polarimetry. This polarimeter was built following the design of  \citet{magalhaes} and uses an achromatic half-wave retarder and a Savart plate as an analyser. The polarimetric measurements and their errors are obtained from a least-squares solution applied to the measurements at different half-wave plate positions. The overall acquisition process and the data reduction pipeline is essentially identical to that previously used in \citet{gil17}.

 The observational circumstances and results of our polarimetric observations are presented in Table 1. For each asteroid, we list the mean time of observations in UT, the adopted filter, the phase angle $\alpha$, the measured polarisation degree $\textit{P}$ and the position angle $\theta$ in the equatorial system, their root-mean-square errors  $\sigma_{\textit{P}}$ and  $\sigma_{\theta}$, the polarisation degree $\textit{P}_{r,}$ and the position angle  $\theta_{r}$ in the coordinate system referring to the scattering plane, as defined by \citet{zellner}. 
  
\begin{table*}
\centering
    \caption[]{Observational circumstances and results of our polarimetric observations.}
\begin{tabular}{lccrllrrrr}
\hline
 \hhline{~~~}\\
Asteroid          & Date, UT        & Filter &$\alpha$, deg & $\textit{P}$, \% & $\sigma_{\textit{P}}$, \% & $\theta\,\,\,$, deg    &$\sigma_{\theta}$, deg & $\textit{P}_{r}$, \% & $\theta_{r}$, deg \\
 \hhline{~~~}\\
 \hline
   \hhline{~~~}\\
(22) Kalliope$^1$    & 2021   11 08.90 & R      & 18.7   & 0.315 & 0.049  & 92.8  & 4.5     & -0.315  & 90.0    \\
(69) Hesperia$^2$    & 2018 08 07.34   & V      & 17.3   & 0.46  & 0.08   & 59.4  & 2.2     & -0.45   & 84.1    \\
(77) Frigga$^1$      & 2020   11 17.13 & R      & 23.3   & 0.179 & 0.045  & 19.2  & 7.2     & 0.179   & 177.9   \\
(129) Antigone$^1$   & 2021   11 08.99 & R      & 7.97   & 0.970 & 0.035  & 36.3  & 1.0     & -0.967  & 92.2    \\
(135)   Hertha    & 2021   03 04.35 & B      & 15.9   & 0.603 & 0.037  & 97.4  & 1.7     & -0.600  & 87.1    \\
(135) Hertha      & 2021   03 04.35 & V      & 15.9   & 0.734 & 0.052  & 98.7  & 2.0     & -0.733  & 88.4    \\
(135) Hertha      & 2021   03 04.35 & R      & 15.9   & 0.641 & 0.031  & 96.5  & 1.4     & -0.635  & 86.2    \\
(135) Hertha      & 2021   03 18.39 & B      & 18.4   & 0.305 & 0.085  & 94.6  & 7.8     & -0.300  & 85      \\
(135) Hertha      & 2021   03 18.39 & V      & 18.4   & 0.638 & 0.071  & 92.3  & 3.2     & -0.617  & 82.7    \\
(135) Hertha      & 2021   03 18.39 & R      & 18.4   & 0.488 & 0.053  & 96.5  & 3.1     & -0.485  & 86.9    \\
(184)   Dejopeja  & 2021   03 15.61 & B      & 18.8   & 0.268 & 0.080  & 63.5  & 8.4     & -0.119  & 58.2    \\
(184)   Dejopeja  & 2021   03 15.61 & V      & 18.8   & 0.507 & 0.113  & 73.2  & 6.3     & -0.363  & 67.9    \\
(184)   Dejopeja  & 2021   03 15.61 & R      & 18.8   & 0.552 & 0.043  & 60.8  & 2.2     & -0.198  & 55.5    \\
(184)   Dejopeja  & 2021   03 20.61 & B      & 18.6   & 0.240 & 0.078  & 96.1  & 8.4     & -0.108  & 58.4    \\
(184)   Dejopeja  & 2021   03 20.61 & V      & 18.6   & 0.548 & 0.117  & 84.4  & 6.0     & -0.510  & 79.3    \\
(184)   Dejopeja  & 2021   03 20.61 & R      & 18.6   & 0.324 & 0.084  & 99.6  & 7.3     & -0.320  & 94.5    \\
(201) Penelope$^2$   & 2018 05 18.05   & V      & 17.1   & 0.32  & 0.06   & 108.6 & 3.0     & -0.31   & 83.0    \\
(224) Oceana$^2$     & 2021 08 07.28   & V      & 15.8   & 0.58  & 0.19   & 59.0  & 9.2     & -0.55   & 81.0    \\
(250) Bettina     & 2021   03 19.26 & B      & 20.8   & 0.080 & 0.042  & 39.0  & 13.9    & -0.013  & 49.7    \\
(250) Bettina     & 2021   03 19.26 & V      & 20.8   & 0.171 & 0.051  & 62.1  & 8.2     & -0.141  & 72.8    \\
(250) Bettina     & 2021   03 19.26 & R      & 20.8   & 0.146 & 0.032  & 75.0  & 6.2     & -0.144  & 85.7    \\
(337) Devosa$^1$     & 2021   11 08.94 & R      & 5.9    & 1.079 & 0.031  & 8.0   & 1.0     & -1.079  & 90.8    \\
(347) Pariana     & 2021   03 15.58 & B      & 24.8   & 0.354 & 0.049  & 175.5 & 4.0     & 0.336   & 170.9   \\
(347) Pariana     & 2021   03 15.58 & V      & 24.8   & 0.487 & 0.098  & 7.8   & 5.7     & 0.484   & 3.2     \\
(347) Pariana     & 2021   03 15.58 & R      & 24.8   & 0.416 & 0.057  & 0.7   & 3.9     & 0.412   & 176.1   \\
(347) Pariana     & 2021   03 19.59 & B      & 24.5   & 0.437 & 0.070  & 1.6   & 4.6     & 0.436   & 177.8   \\
(347) Pariana     & 2021   03 19.59 & V      & 24.5   & 0.336 & 0.074  & 165.9 & 6.2     & 0.273   & 162.1   \\
(347) Pariana     & 2021   03 19.59 & R      & 24.5   & 0.379 & 0.045  & 176.9 & 3.4     & 0.368   & 173.1   \\
(369) Aeria$^2$      & 2018 05 18.26   & V      & 11.2   & 0.85  & 0.07   & 73.6  & 1.9     & -0.82   & 82.4    \\
(369) Aeria$^1$      & 2020   11 17.15 & R      & 20.9   & 0.052 & 0.055  & 121.1 & 30.3    & -0.049  & 100.4   \\
(382) Dodona      & 2021   03 18.45 & B      & 11.5   & 0.738 & 0.082  & 98.0  & 3.2     & -0.725  & 84.6    \\
(382) Dodona      & 2021   03 18.45 & V      & 11.5   & 0.979 & 0.105  & 102.2 & 3.2     & -0.978  & 88.8    \\
(382) Dodona      & 2021   03 18.45 & R      & 11.5   & 1.017 & 0.075  & 102.3 & 2.1     & -1.016  & 88.9    \\
(441) Bathilde$^1$   & 2021   11 05.09 & R      & 20.9   & 0.137 & 0.323  & 81.3  & 67.5    & -0.080  & 62.9    \\
(441) Bathilde$^1$   & 2021 11 09.15   & R      & 21.2   & 0.248 & 0.093  & 115.9 & 10.7    & -0.241  & 97.0    \\
(678) Fredegundis & 2021   03 04.52 & B      & 6.8    & 0.810 & 0.050  & 136.9 & 1.8     & -0.809  & 88.5    \\
(678) Fredegundis & 2021   03 04.52 & V      & 6.8    & 0.968 & 0.067  & 137.7 & 2.0     & -0.968  & 89.3    \\
(678) Fredegundis & 2021   03 04.52 & R      & 6.8    & 0.881 & 0.042  & 135.3 & 1.4     & -0.876  & 86.9    \\
(741) Botolphia   & 2021   03 15.50 & B      & 8.7    & 0.907 & 0.099  & 77.1  & 3.1     & -0.907  & 89.1    \\
(741) Botolphia   & 2021   03 15.50 & V      & 8.7    & 1.259 & 0.166  & 76.6  & 3.8     & -1.257  & 88.6    \\
(741) Botolphia   & 2021   03 15.50 & R      & 8.7    & 1.129 & 0.101  & 73.5  & 2.6     & -1.115  & 85.5    \\
(741) Botolphia   & 2021   03 22.44 & B      & 6.5    & 0.962 & 0.055  & 60.1  & 1.6     & -0.962  & 90.1    \\
(741) Botolphia   & 2021   03 22.44 & V      & 6.5    & 1.003 & 0.058  & 57.1  & 1.7     & -0.998  & 87.1    \\
(741) Botolphia   & 2021   03 22.44 & R      & 6.5    & 0.967 & 0.041  & 59.7  & 1.2     & -0.967  & 89.7    \\
(755) Quintilla   & 2021   05 08.34 & B      & 8.7    & 0.997 & 0.048  & 113.8 & 1.4     & -0.987  & 86.0      \\
(755) Quintilla   & 2021   05 08.34 & V      & 8.7    & 1.043 & 0.069  & 114.8 & 1.9     & -1.037  & 87.0      \\
(755) Quintilla   & 2021   05 08.34 & R      & 8.7    & 1.155 & 0.034  & 115.0 & 0.8     & -1.149  & 87.2    \\
(755) Quintilla   & 2021   05 13.34 & B      & 10.6   & 1.070 & 0.061  & 114.1 & 1.6     & -1.066  & 87.4    \\
(755) Quintilla   & 2021   05 13.34 & V      & 10.6   & 1.186 & 0.080  & 116.5 & 1.9     & -1.186  & 89.8    \\
(755) Quintilla   & 2021   05 13.34 & R      & 10.6   & 1.141 & 0.053  & 115.2 & 1.3     & -1.139  & 88.5    \\
(755) Quintilla   & 2021   06 01.37 & B      & 16.3   & 0.869 & 0.072  & 112.7 & 2.4     & -0.868  & 88.4    \\
(755) Quintilla   & 2021   06 01.37 & V      & 16.3   & 0.676 & 0.108  & 109.1 & 4.5     & -0.665  & 84.8    \\
(755) Quintilla   & 2021   06 01.37 & R      & 16.3   & 0.588 & 0.056  & 110.1 & 2.7     & -0.582  & 85.8    \\
(758) Mancunia    & 2021   05 17.40 & B      & 8.3    & 0.905 & 0.070  & 81.5  & 2.2     & -0.897  & 86.2    \\
(758) Mancunia    & 2021   05 17.40 & V      & 8.3    & 1.307 & 0.130  & 85.9  & 2.8     & -1.307  & 90.6    \\
(758) Mancunia    & 2021   05 17.40 & R      & 8.3    & 1.183 & 0.061  & 82.1  & 1.5     & -1.176  & 86.8    \\
(758) Mancunia    & 2021   05 18.42 & B      & 8.1    & 1.203 & 0.064  & 83.7  & 1.5     & -1.201  & 88.5    \\
(758) Mancunia    & 2021   05 18.42 & V      & 8.1    & 1.039 & 0.097  & 82.6  & 2.7     & -1.035  & 87.4    \\
\end{tabular}
\end{table*}
\begin{table*}
\centering
\begin{tabular}{llcrllrrrr}

\multicolumn{10}{l}{Continuation of Table 1.\, Observational circumstances and results of polarimetric observations.}\\
\hhline{~~~}   \\
\hline\\
 \hhline{~~~}\\
Asteroid          & Date, UT        & Filter &$\alpha$, deg & $\textit{P}$, \% & $\sigma_{\textit{P}}$, \% & $\theta\,\,\,$    &$\sigma_{\theta}$, deg & $\textit{P}_{r}$, \% & $\theta_{r}$, deg \\
 \hhline{~~~}\\
 \hline
\hhline{~~~}\\
(758) Mancunia    & 2021   05 18.42 & R      & 8.1    & 1.195 & 0.067  & 81.6  & 1.6     & -1.186  & 86.4    \\
(779) Nina$^1$       & 2021 11 08.78   & R      & 10.7   & 1.026 & 0.043  & 49.2  & 1.2     & -1.021  & 92.8    \\
(785) Zwetana     & 2021   03 19.31 & B      & 24.0   & 0.256 & 0.077  & 114.3 & 8.3     & -0.247  & 97.8    \\
(785) Zwetana     & 2021   03 19.31 & V      & 24.0   & 0.054 & 0.084  & 1.0   & 28.8    & 0.046   & 164.5   \\
(785) Zwetana     & 2021   03 19.31 & R      & 24.0   & 0.182 & 0.058  & 15.4  & 8.9     & 0.182   & 178.9   \\
(872) Holda       & 2021   05 08.45 & B      & 11.3   & 0.991 & 0.050  & 74.9  & 1.5     & -0.979  & 85.5    \\
(872) Holda       & 2021   05 08.45 & V      & 11.3   & 0.995 & 0.059  & 76.2  & 1.7     & -0.989  & 86.8    \\
(872) Holda       & 2021   05 08.45 & R      & 11.3   & 1.033 & 0.041  & 77.0  & 1.2     & -1.029  & 87.6    \\
(872) Holda       & 2021   05 13.44 & B      & 9.4    & 1.053 & 0.052  & 68.7  & 1.4     & -1.030  & 84.0      \\
(872) Holda       & 2021   05 13.44 & V      & 9.4    & 0.970 & 0.058  & 73.6  & 1.7     & -0.969  & 88.9    \\
(872) Holda       & 2021   05 13.44 & R      & 9.4    & 1.238 & 0.040  & 70.7  & 0.9     & -1.226  & 86.0      \\
(872) Holda       & 2021   05 15.42 & B      & 8.7    & 1.132 & 0.054  & 71.1  & 1.4     & -1.131  & 88.8    \\
(872) Holda       & 2021   05 15.42 & V      & 8.7    & 1.110 & 0.100  & 71.5  & 2.6     & -1.110  & 89.2    \\
(872) Holda       & 2021   05 15.42 & R      & 8.7    & 1.081 & 0.049  & 72.6  & 1.3     & -1.081  & 90.3    \\
(1222) Tina$^1$      & 2021   11 04.75 & R      & 23.12  & 0.875 & 0.158  & 171.2 & 5.2     & 0.873   & 2.1     \\
(1222) Tina$^1$      & 2021   11 08.70 & R      & 23.52  & 0.810 & 0.053  & 174.2 & 1.9     & 0.786   & 6.9  \\
 &                 &        &        &       &        &       &         &         &         \\
%\multicolumn{10}{| l |}{End of Table}\\

\hline
  \noalign{\smallskip}

$^1${\footnotesize observations at Rozhen} & \\
$^2${\footnotesize observations at CASLEO} &
\end{tabular}
\end{table*}

\subsection{Wavelength dependence}
Simultaneous BVR measurements obtained for 11 asteroids provide an opportunity to estimate the wavelength dependence of the polarisation degree. The measured polarisation degrees versus phase angle in BVR filters for all observed asteroids are shown in Fig.1a. We found that the differences between measurements in different filters are small and typically fall within the errors of the measurements. To evaluate wavelength dependence we calculated the linear slope of the polarisation degree in the wavelength range of 0.44-0.64 $\mu$m for each observed asteroid. These slopes are plotted in Fig.1b as a function of the phase angle at which they were measured. Observed asteroids typically show a small negative slope, which means that the negative polarisation branch is deeper with increasing wavelength. This trend is consistent with previous findings for M-type asteroids in the wider wavelength range (\citealt{belskaya}). For two asteroids, (755) Quintilla and (785) Zwetana, measurements at the phase angle $\alpha$ \textgreater 15\textdegree\, revealed an opposite trend, and whether this trend is real or not will require further verification.  
   
   \begin{figure*}[ht!]
        % \vspace*{-1.0 cm}
        \begin{center}
        %[width=\hsize] =6.4in
                \includegraphics[width=6.4in]{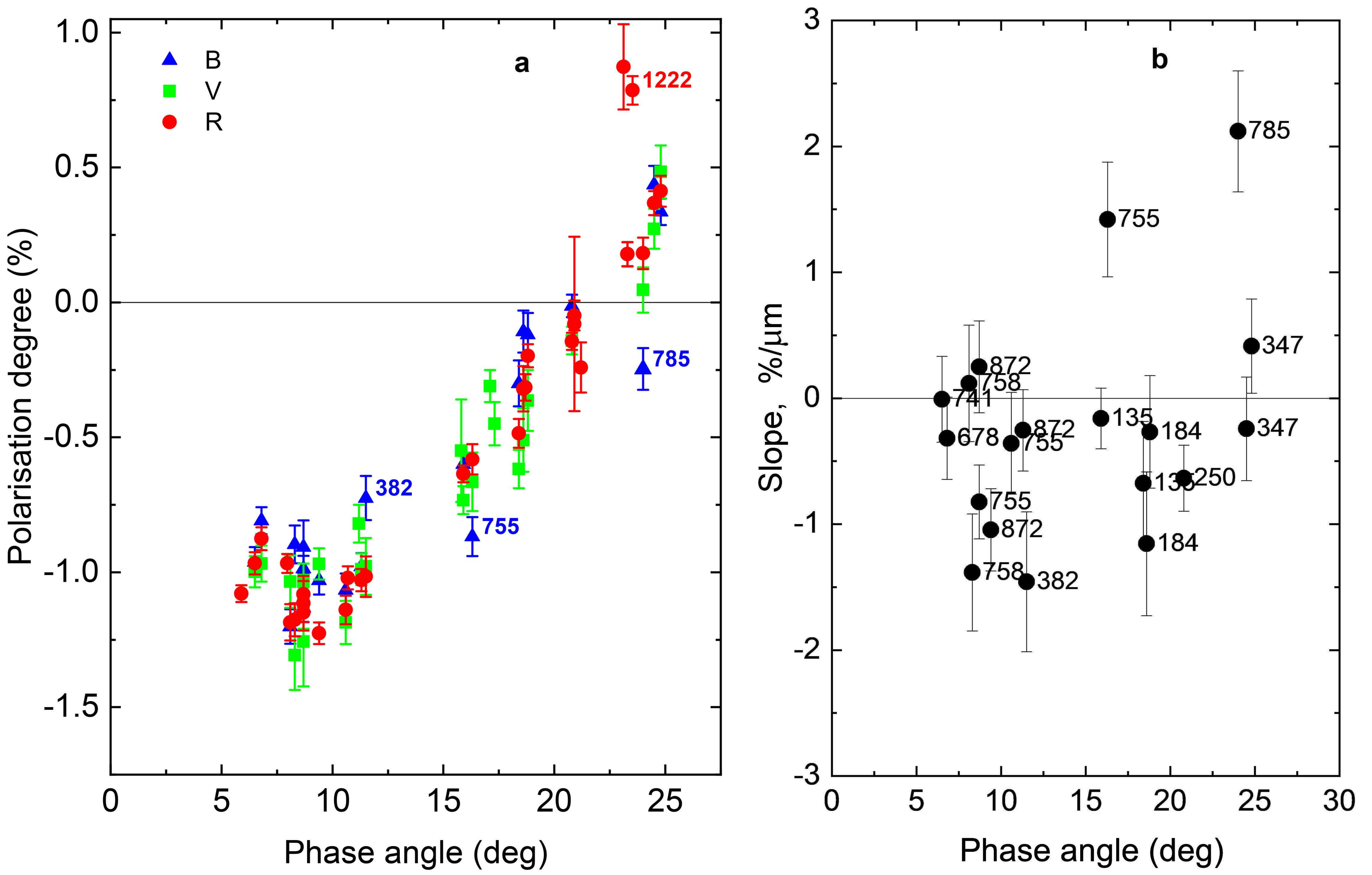} 
                % \vspace*{-1.0 cm}
                \caption{Polarisation degree in the B, V, R filters (a) and polarisation spectral slope (b) as a function of phase angle for the observed asteroids. Asteroids with large deviations of the measured polarisation degree from typical values are marked with their numbers in Fig.1a.}
                \label{fig1}
        \end{center}
\end{figure*}
   
 \section{Determination of polarimetric parameters }  
 
We analysed the new observational data together with the available literature data on M/X-type asteroids. The published data and all relevant references are given in the catalogues of asteroid polarimetry (\citealt{lupishko},  \citealt{gil}).  In case of a discrepancy between the data obtained by different authors, we carefully examined original publications. Individual polarimetric phase curves of the observed asteroids including both new and already published data are shown in Fig.2. For some asteroids, we combined observations in V and R filters, since the expected differences of the polarisation degree measured in these filters are within the accuracy of observational data. The filter in which observations were obtained is indicated in each figure. 
 For fitting the data, we used the exponential-linear function (\citealt{kaasa}; \citealt{muinonen}) in the following form:
$$
\textit{P}_\mathrm{r} = A(e^{-\alpha/B}-1) + C \cdot\alpha,
$$
where $\alpha$ is the phase angle expressed in degrees, and  \textit{A}, \textit{B}, and \textit{C} are free parameters. The function is known to work well up to $\alpha\leqslant$ 30\textdegree\, (\citealt{muinonen}).  We used the Levenberg-Marquardt algorithm to calculate the best-fitting parameters \textit{A}, \textit{B,} and \textit{C,} and their errors were obtained using a Monte Carlo simulation. With these parameters, it is possible to calculate the usual polarimetric parameters and estimate their uncertainties by a propagation of errors. The parameters $\textit{P}_{min}$ and inversion angle $\alpha_{inv}$ were found using the procedure described in the paper by Muinonen et al. (2009). For objects with few measurements, we fixed the parameter C within the range of 0.1--0.15. These values of polarimetric slope parameter are inherent for moderate albedo asteroids with geometric albedos ${p}_{V}\sim 0.1-0.2$ according to empirical polarimetric slope-albedo relationships (\citealt{cellino},  \citealt{lupishko18}). In the case of a single available measurement in the vicinity of $\textit{P}_{min}$ (at phase angles of 7-11\textdegree), we preferred to use the measured value as $\textit{P}_{min}$ rather than the value derived from the fitting curve. The inversion angle was determined only when the measurements were available at phase angles close to the inversion angle ($\alpha$ $\geqslant{19}$\textdegree). We would like to mention that different ways of determining the polarimetric parameters give consistent results within the assigned uncertainties and do not affect our further conclusions.

 \begin{figure*}[ht!]
        \begin{center}
        \includegraphics[width=\hsize]{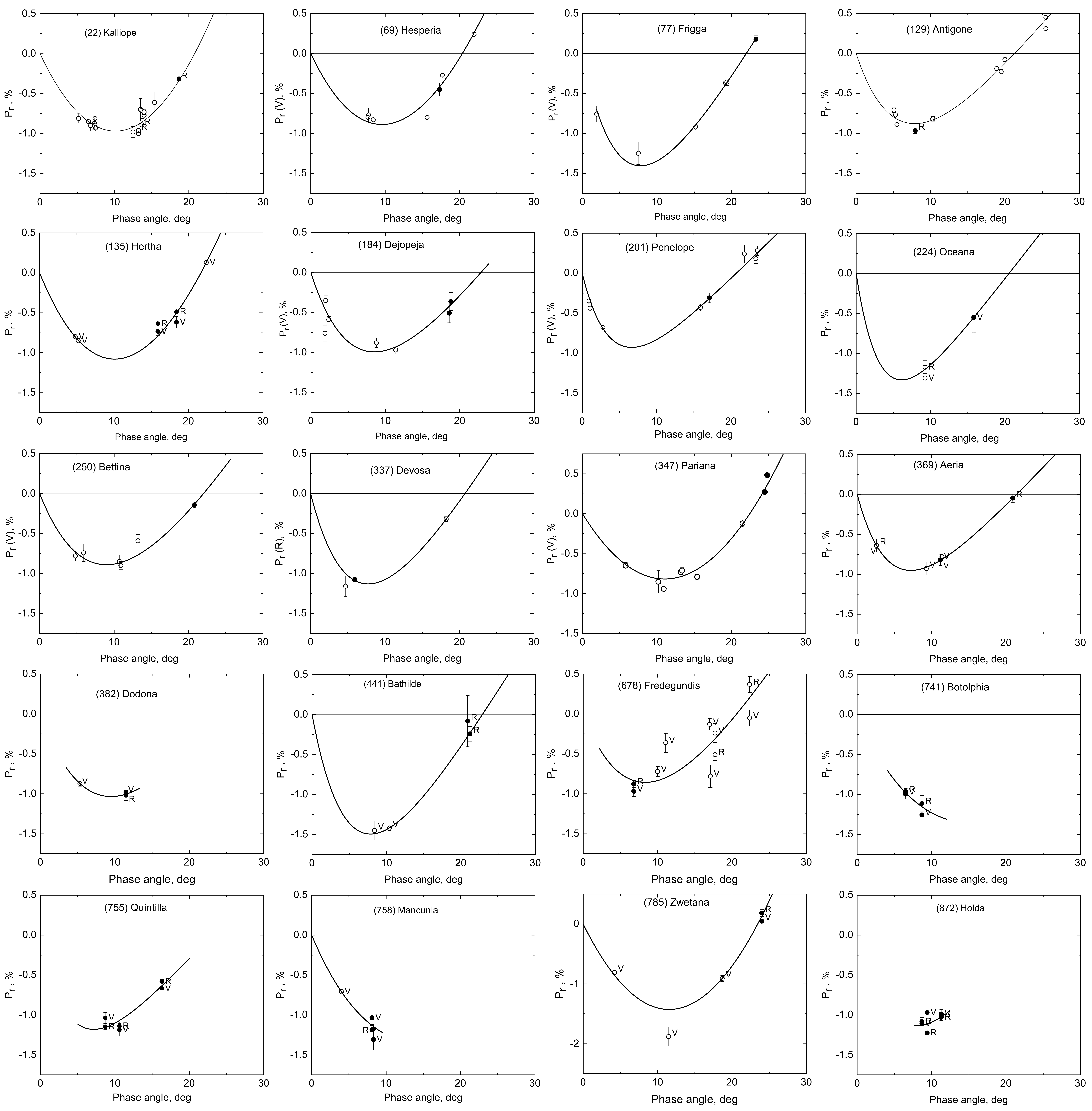} 
        \caption{Polarisation-phase curves of the measured asteroids. New observations are shown by filled symbols and previously published data are shown by open symbols. The line shows the best fit of the exponential-linear function (\citealt{kaasa}, \citealt{muinonen}).}
        \label{fig2}
        \end{center}
\end{figure*}
  
  The best-fitted curves are plotted in Fig.2. Data for the asteroids (779) Nina and (1222) Tina are not presented in Fig.2 since they have only two points each. Typically, the deviation of observational data from the fitting function is within the errors of measurements. Only two asteroids from our sample, (678) Fredegundis and (785) Zwetana, show a large scatter of data points (see Fig.2). Such a scatter is difficult to explain by possible variations of polarisation degree over the surface, since the expected variations for main belt asteroids are small (e.g. \citealt{cellino16}). There are most likely technical problems with some measurements. 
  
  With our new observations, at least one of two polarimetric parameters $\textit{P}_{min}$ or $\alpha_{inv}$ was estimated for the first time or revised for all considered asteroids. The determined polarimetric parameters of these asteroids (marked in bold) are given in Table 2. We also included 19 asteroids of M/X-type for which polarimetric parameters can be determined based on previously published data available in catalogues of asteroid polarimetry (Lupishko 2019, Gil-Hutton 2017). The observations in V or R filters of each asteroid were fitted by the described above fitting procedure to determine $\textit{P}_{min}$ and $\alpha_{inv}$. Thus, we obtained a homogeneous data set of polarimetric parameters for 41 M/X-type asteroids presented in Table 2. Table 2 also contains their types according to the \citet{tholen} and Bus-DeMeo classification schemes (\citealt{bus},  \citealt{demeo}), albedos from Akari data (\citealt{ali}) and WISE data (\citealt{mainzer}), and the spectral slopes in the 1.7-2.4 $\mu$m wavelength range. The spectral slopes were calculated from the available near-infrared spectra of these asteroids obtained by \citet{clark}, \citet{fornasier}, Ockert-Bell et al. (\citeyear{ockert08}, \citeyear{ockert10}), \citet{hardersen11}, and \citet{neeley} (see subsection 4.4 for details).
 \begin{table*}
 \centering
  \caption[]{Polarimetric parameters and other characteristics of M/X-type asteroids.}
\begin{tabular}{rlcccccc}
\hline
 \hhline{~~~}\\
 \multicolumn{2}{c}{Asteroid$^1$} & Type   & $p_{V}$  & $p_{V}$ & $|\textit{P}_{min}|$ & $\alpha_{inv}$ & NIR slope$^2$       \\
                  &                 & (Tholen / Bus-DeMeo) &(Akari)   &  (WISE)  &   & &                \\

  \hline
  \hhline{~~~}\\                 
16              & Psyche      & M\,/\,Xk    & 0.19 $\pm{0.08}$             & -                                      & 1.06 $\pm{0.05}$ & 22.3 $\pm{0.6}$    & 0.17 $\pm{0.07}$            \\
21              & Lutetia     & M\,/\,Xc    & 0.15 $\pm{0.03}$              & -                 & 1.32 $\pm{0.07}$    & 25.0 $\pm{0.4}$   & -0.03 $\pm{0.02}$                                                    \\
 \textbf{22}      &  \textbf{Kalliope}    & M\,/\,X       & 0.22 $\pm{0.04}$   & 0.17 $\pm{0.01}$       & 0.97 $\pm{0.07}$    & 21.0 $\pm{1.0}$    & 0.16 $\pm{0.02}$   \\
55    & Pandora     & M\,/\,Xk     & 0.33 $\pm{0.07}$           & -          & 0.97 $\pm{0.06}$    & 20.6 $\pm{0.8}$     & 0.15 $\pm{0.05}$      \\
\textbf{69}     & \textbf{Hesperia}    & M\,/ Xk    & 0.12 $\pm{0.03}$  & -         & 0.85 $\pm{0.10}$    & 20.5 $\pm{0.8}$   & 0.12 $\pm{0.04}$        \\
75                     & Eurydike    & M\,/ Xk     & 0.11 $\pm{0.02}$   & 0.12 $\pm{0.03}$        & -    & 20.1 $\pm{0.8}$    & -            \\
\textbf{77}            & \textbf{Frigga}      & MU\,/\,Xe   & 0.13 $\pm{0.03}$   & 0.23 $\pm{0.03}$      & 1.30 $\pm{0.15}$  & 22.0 $\pm{0.5}$ & 0.07 $\pm{0.03}$ \\
92                 & Undina      & X\,/\,Xk   & 0.25 $\pm{0.05}$ & -     & 0.80 $\pm{0.15}$    & -        & -       \\
97         & Klotho      & M\,/ Xc      & 0.18 $\pm{0.04}$    & 0.25 $\pm{0.04}$       & 1.35 $\pm{0.15}$    & 22.6 $\pm{0.9}$    & -0.03 $\pm{0.03}$   \\
110     & Lydia       & M\,/ Xk     & 0.18 $\pm{0.04}$     & 0.17 $\pm{0.02}$   & 0.90 $\pm{0.10}$     & -        &0.13 $\pm{0.03}$       \\
125  & Liberatrix  & M\,/ X\,      & 0.29 $\pm{0.06}$  & 0.18 $\pm{0.04}$   & 0.85 $\pm{0.15}$    & 20.4 $\pm{0.9}$     & 0.20 $\pm{0.06}$        \\
\textbf{129}     & \textbf{Antigone}    & M\,/ X\,      & 0.17 $\pm{0.03}$   & 0.15 $\pm{0.06}$           & 0.90 $\pm{0.10}$    & 21.4 $\pm{0.6}$   & 0.18 $\pm{0.05}$     \\
132  & Aethra      & M\,/ Xe      & 0.17 $\pm{0.03}$   & -     & 1.10 $\pm{0.15}$    & 19.5 $\pm{0.5}$      & -       \\
\textbf{135}     & \textbf{Hertha}      & M\,/ Xk    & 0.17 $\pm{0.03}$   & 0.18 $\pm{0.03}$        & 0.95 $\pm{0.10}$     & 21.6 $\pm{0.8}$ & 0.13 $\pm{0.03}$    \\
161    & Athor       & M\,/ Xc    & 0.23 $\pm{0.05}$      & -    & -       & 18.9 $\pm{1.0}$  & -  \\
\textbf{184}  & \textbf{Dejopeja}    & X\,/ X\,       & 0.17 $\pm{0.03}$ & 0.22 $\pm{0.02}$ & 1.00 $\pm{0.10}$    & -        & 0.17 $\pm{0.05}$         \\
\textbf{201}      & \textbf{Penelope}    & M\,/ Xk       & 0.19 $\pm{0.04}$ & 0.10 $\pm{0.03}$           & -          & 20.9 $\pm{0.8}$   & -       \\
216      & Kleopatra   & M\,/ Xe     & 0.17 $\pm{0.03}$  & 0.15 $\pm{0.03}$    & 1.00 $\pm{0.10}$    & 21.2 $\pm{1.0}$  & 0.16 $\pm{0.02}$ \\
\textbf{224}     & \textbf{Oceana}       & M\,/ X,T$^3$  & 0.19 $\pm{0.04}$ & 0.24 $\pm{0.05}$    & 1.20 $\pm{0.15}$    & -  & 0.05 $\pm{0.01}$   \\
\textbf{250}     & \textbf{Bettina}      & M\,/ Xk     & 0.14 $\pm{0.03}$ & 0.11 $\pm{0.02}$     & 0.90 $\pm{0.12}$   & 22.0 $\pm{0.9}$ & 0.21 $\pm{0.02}$   \\
325        & Heidelberga & M\,/  -\,\,     & 0.10 $\pm{0.02}$  & -     & 0.95 $\pm{0.15}$    & -        & 0.17 $\pm{0.05}$        \\
\textbf{337}     & \textbf{Devosa}       & X\,/ Xk     & 0.12 $\pm{0.02}$ & - & 1.15 $\pm{0.12}$    & -  & 0.12 $\pm{0.03}$  \\
338    & Budrosa     & M\,/ Xk     & 0.14 $\pm{0.03}$  & 0.28 $\pm{0.05}$   & 0.98 $\pm{0.10}$    & -     & 0.12 $\pm{0.01}$  \\
\textbf{347}      & \textbf{Pariana}      & M\,/ -\,\,   & 0.17 $\pm{0.03}$  & 0.22 $\pm{0.05}$     & 0.95 $\pm{0.15}$  & 22.2 $\pm{0.8}$ & 0.18 $\pm{0.03}$  \\
359     & Georgia     & CXM\,/ Xk       & 0.13 $\pm{0.03}$   & -    & -      & 20.8 $\pm{0.6}$  & -      \\
\textbf{369}     & \textbf{Aeria}        & M\,/ -\,\,        &0.15 $\pm{0.03}$  & 0.17 $\pm{0.04}$  & 0.95 $\pm{0.10}$  & 21.1 $\pm{0.9}$ & 0.19 $\pm{0.08}$  \\
\textbf{382}     & \textbf{Dodona}       & M\,/ -\,\,        & 0.19 $\pm{0.04}$ &0.13 $\pm{0.02}$ & 1.03 $\pm{0.10}$  & -  & 0.14 $\pm{0.03}$        \\
\textbf{441}    & \textbf{Bathilde}     & M\,/ Xk        & 0.16 $\pm{0.03}$ &  0.20 $\pm{0.01}$   & 1.40 $\pm{0.10}$    & 22.8 $\pm{1.0}$  & 0.04 $\pm{0.03}$   \\
504   & Cora        & -\,/ X      & 0.21 $\pm{0.04}$  & 0.34 $\pm{0.05}$     & 1.00 $\pm{0.10}$    & -       & -          \\
558  & Carmen      & M\,/ Xk$^3$  & 0.14 $\pm{0.03}$ & 0.13 $\pm{0.04}$ & 0.80 $\pm{0.06}$    & -       & 0.20 $\pm{0.09}$         \\
\textbf{678} & \textbf{Fredegundis} & -\,/ X       & 0.19 $\pm{0.04}$  & 0.34 $\pm{0.06}$     & 0.90 $\pm{0.08}$    & 20.6 $\pm{1.0}$ & 0.12 $\pm{0.02}$     \\
\textbf{741}  & \textbf{Botolphia}   & X\,/ X       & 0.20 $\pm{0.04}$  & -  & 1.15 $\pm{0.10}$    & -        & -    \\
\textbf{755}   & \textbf{Quintilla}   & M\,/ -\,\,        & 0.22 $\pm{0.09}$  &0.12 $\pm{0.01}$    & 1.15 $\pm{0.10}$    & -        & 0.04 $\pm{0.08}$     \\
757      & Portlandia  & XF\,/ Xk        & 0.13 $\pm{0.03}$ & 0.22 $\pm{0.03}$  & 1.00 $\pm{0.20}$    & 16.6 $\pm{0.7}$  & 0.05 $\pm{0.03}$ \\
\textbf{758}      & \textbf{Mancunia}    & X\,/ -\,\,         & 0.11 $\pm{0.02}$  & 0.12 $\pm{0.02}$     & 1.25 $\pm{0.10}$    & -  & 0.08 $\pm{0.03}$    \\
\textbf{779}  & \textbf{Nina}        & -\,/ X   & 0.14 $\pm{0.03}$ & 0.16 $\pm{0.02}$      & 1.00 $\pm{0.10}$     & -         & -       \\
\textbf{785}      & \textbf{Zwetana}     & M\,/ Cb       & 0.13 $\pm{0.03}$ & -  & 1.88 $\pm{0.20}$    & 24.0 $\pm{0.6}$  &  0.21 $\pm{0.02}$     \\
796         & Sarita      & XD\,/ X      & 0.23 $\pm{0.05}$   & 0.21 $\pm{0.03}$   & 1.00 $\pm{0.10}$    & 21.0 $\pm{1.0}$  &  0.20 $\pm{0.08}$   \\
849     & Ara         & M\,/ -\,\,       & 0.30 $\pm{0.06}$ & 0.13 $\pm{0.04}$   & 0.95 $\pm{0.10}$    & -        & 0.21 $\pm{0.08}$               \\
\textbf{872}  & \textbf{Holda}       & M\,/ X\,        & 0.17 $\pm{0.03}$ & 0.24 $\pm{0.03}$  & 1.13 $\pm{0.10}$ & -        & 0.13 $\pm{0.02}$    \\
\textbf{1222}   & \textbf{Tina}        & -\,/ X       & 0.09 $\pm{0.02}$   & 0.20 $\pm{0.05}$   & -   & $\sim$20 & -    \\
&  \\
 \hline
 \noalign{\smallskip}
 \multicolumn{8}{l}{$^1${\footnotesize\,Asteroids observed in the present work are marked in bold.} }  \\ 
 \multicolumn{8}{l}{$^2${\footnotesize\,The spectral slope in the 1.7-2.4 $\mu$m wavelength range calculated using the published data by \citet{clark}, \citet{fornasier},} }  \\
 \multicolumn{8}{l}{\footnotesize  \,\,\,\,Ockert-Bell et al. (\citeyear{ockert08}, \citeyear{ockert10}), \citet{hardersen11}, and \citet{neeley}.}   \\
 \multicolumn{3}{l}{$^3${\footnotesize\,Type from \citet{lazzaro}.}} \\ 
\end{tabular}
\end{table*}

In total, 41 asteroids are listed in Table 2. The data set includes 29 asteroids classified as M types in Tholen’s classification (\citealt{tholen}), which is about 70\% of all asteroids classified as M types, and 12 asteroids classified as X types in the available classifications ( \citealt{tholen}, \citealt{bus}, \citealt{lazzaro}, \citealt{demeo}). 

 \section{Search for relationships}  
  
We used the data set of polarimetric parameters of M-type asteroids (Table 2) to search for relationships between the parameters characterising their properties obtained by various techniques.

 \subsection{$\textit{P}_{min}$ versus inversion angle}
   
Table 2 contains 20 asteroids for which both polarimetric parameters $\textit{P}_{min}$ and $\alpha_{inv}$ were determined. The relationship between these two parameters characterising the negative polarisation branch is shown in Fig.3.  For comparison, we plotted the data for other composition types based on the catalogue of polarimetric parameters from \citet{gil}. The polarimetric parameters were calculated in the same way, by fitting the observational data with the exponential-linear function.
   
      \begin{figure}[!h]
        % \vspace*{-1.0 cm}
        \begin{center}
                \includegraphics[width=3.5in]{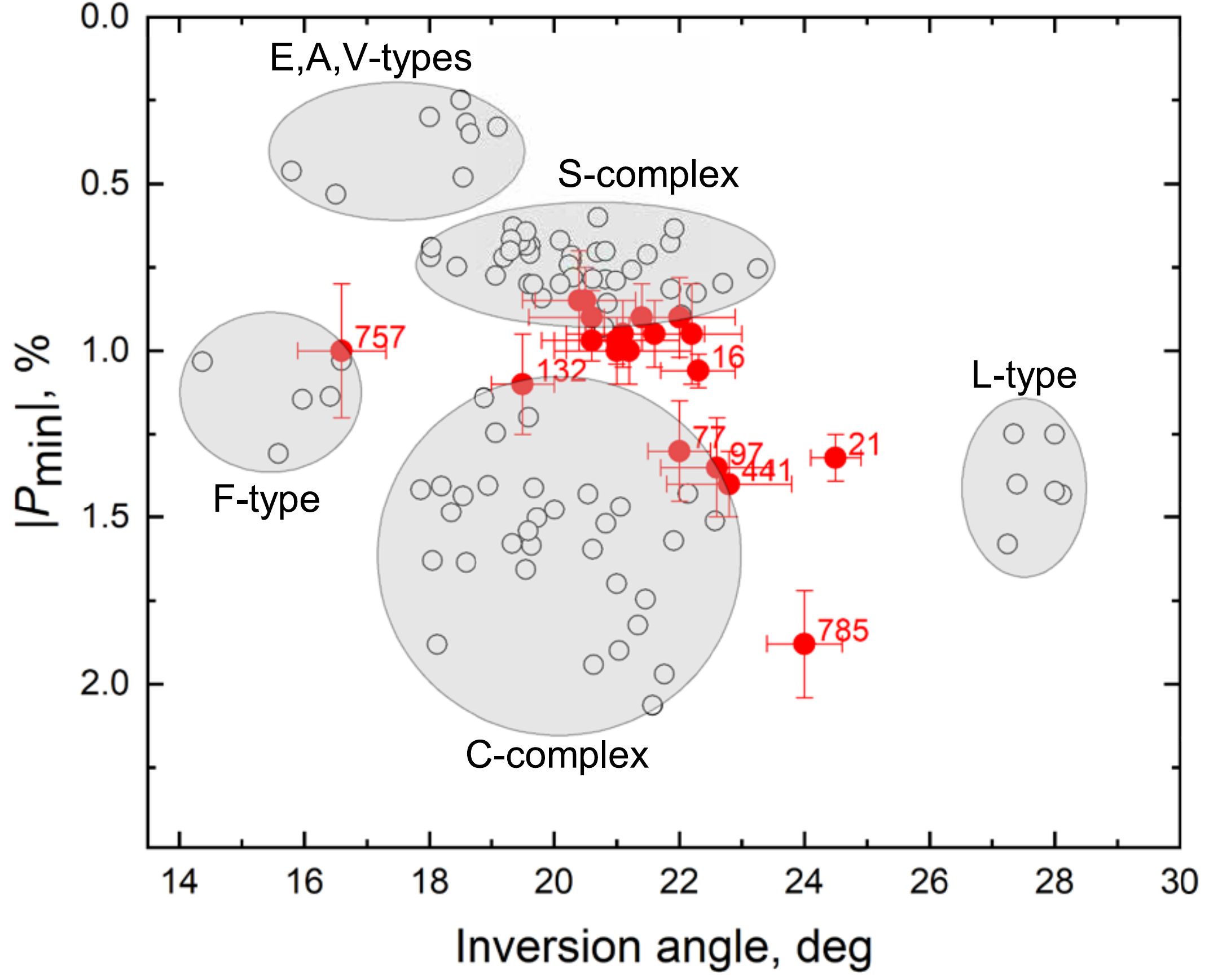} 
                % \vspace*{-1.0 cm}
                \caption{Relationship of $\textit{P}_{min}$ and $\alpha_{inv}$ for asteroids of different composition types. M-type asteroids are shown as filled circles.  }
                \label{fig3}
        \end{center}
\end{figure}

We found that asteroids classified as M types (or moderate albedo X types) have a greater scatter of polarimetric parameters compared to other types (Fig.3). M-type asteroids can be divided into two main sub-groups. The first sub-group of asteroids with $|\textit{P}_{min}|\sim$ 0.9--1\% and $\alpha_{inv}\sim$ 20--22\textdegree\,  is located between the S complex and the C complex and looks like a continuation of the S complex. Several M asteroids fall into the S complex. The second sub-group of asteroids with $|\textit{P}_{min}|\geqslant$ 1.2\% and inversion angle > 22\textdegree\, fall between C-complex and L-type asteroids. 

Two asteroids of our sample, (757) Portlandia and (785) Zwetana, present extreme polarimetric properties. Asteroid (757) Portlandia shows a very small inversion angle typical of low-albedo F-type asteroids (\citealt{belskaya05}). Moreover, this asteroid was initially classified as XF by \citet{tholen}, but its geometric albedo of 0.13 (Akari) or 0.22 (WISE) contradicts this classification. Most likely, the albedo is incorrect and (757) Portlandia belongs to a low-albedo F type. Asteroid (785) Zwetana shows a very deep negative polarisation branch incompatible with its moderate albedo (0.13). 
It was classified as M by \citet{tholen} and Cb by \citet{bus} and \citet{demeo}. Further observations of (785) Zwetana are needed to understand the reasons for its particular properties. 
We do not consider asteroids (757) Portlandia and (785) Zwetana in our search for correlations with radar and spectral data because of their extreme properties, but we include them in our discussion. 

\subsection{$\textit{P}_{min}$ versus geometric albedo}
Albedo is the most important characteristic that distinguishes the M type from the X type. 
Polarimetry can provide independent albedo estimates based on the well-known correlations between polarimetric parameters and albedo (e.g. \citealt{zellner77}). Albedos can be estimated with an uncertainty of about 20\% for $|\textit{P}_{min}|\leqslant{1}$\% (\citealt{cellino}). The deeper negative polarisation branch can occur for asteroids with a wide range of albedos up to 0.2 and cannot be used for reliable albedo estimates of individual asteroids. In our sample, $\textit{P}_{min}$ varies from -0.8 to -1.4\%, which corresponds to albedos in the range of 0.08--0.2 according to the available '$\textit{P}_{min}$ -- albedo' relations (\citealt{cellino}, \citealt{lupishko18}). The range of radiometric albedos for the same sample is broader and span from 0.1 to 0.34 (see Table 2). 
In Fig.4, we plot the dependence of $\textit{P}_{min}$ on radiometric albedos separately for Akari data (\citealt{ali}) and WISE data (\citealt{mainzer}). Two lines correspond to the '$\textit{P}_{min}$ -- albedo' relationship proposed by \citet{lupishko18} and \citet{cellino} to calculate polarimetric albedos. As one can see, radiometric albedos are generally higher than polarimetric albedos and do not show a noticeable trend with $\textit{P}_{min}$. A large scatter of '$\textit{P}_{min}$--radiometric albedo' dependence can be caused both by possible deviations from the relationship and by underestimated errors of radiometric albedos.  This comparison shows that albedos of some asteroids can have significant errors and should be used with caution in their spectral modelling. 

    \begin{figure*}[ht]
        % \vspace*{-1.0 cm}
        \begin{center}
                \includegraphics[width=6.0in]{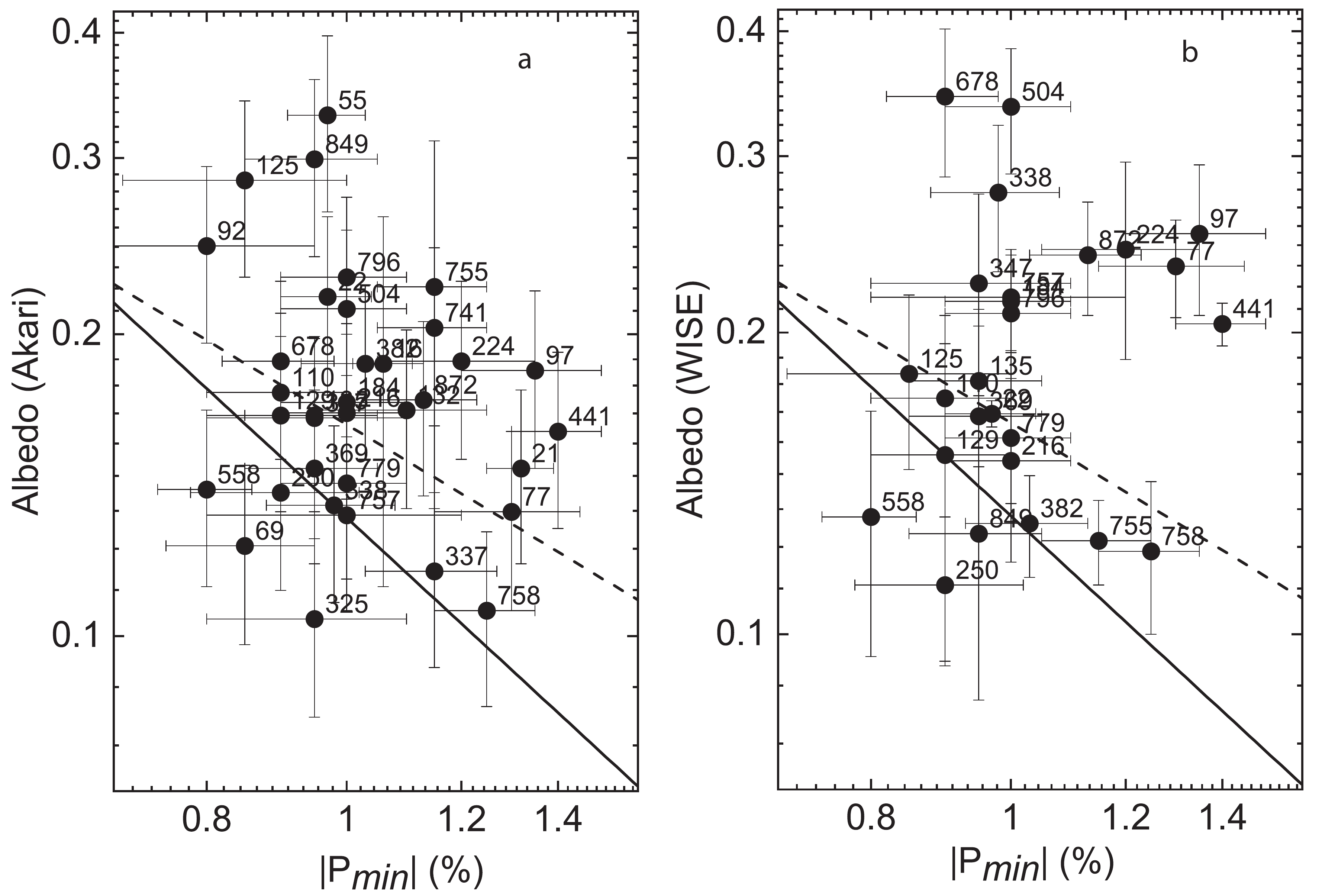} 
                % \vspace*{-1.0 cm}
                \caption{ $\textit{P}_{min}$ versus geometric albedos from infrared data (Akari (a) and WISE (b)).}
                \label{fig4}
        \end{center}
\end{figure*}

\subsection{Polarimetric parameters and radar data}

High radar albedos are considered indicative of a metal-rich surface (\citealt{ostro}, \citealt{shepard08}, \citealt{shepard15}). Polarimetric data are available for 22 of 29 asteroids classified as M/X types for which radar measurements were obtained. Fig.5 shows the relationships of $|\textit{P}_{min}|$ and radar data taken from \citet{shepard15}. The asteroids with lower radar albedos tend to have a deeper negative polarisation branch, with the exception of  asteroid (758) Mancunia (Fig.5a). On the other hand, asteroids with $|\textit{P}_{min}|$ < 1.1\% have diverse radar albedos ranging from low to high values. The correlation of $\textit{P}_{min}$ with radar echo’s circular polarisation ratio $\mu_{c}$ is more evident (Fig.5b). Asteroids that have a deeper negative polarisation branch in optical measurements tend to have a higher circular polarisation ratio $\mu_{c}$. One exception is asteroid (77) Frigga, which is classified as MU in Tholen’s taxonomy (\citealt{tholen}). A strong dependence of circular polarisation ratio on taxonomic type was found by \citet{benner} for near-Earth asteroids. They assumed that different types of asteroids have distinct differences in their near-surface roughness. Two sub-groups seen in the plot of $\textit{P}_{min}$ and radar circular polarisation ratios gave further evidence of diverse taxonomy among M-type asteroids. Both polarimetric parameters $\textit{P}_{min}$ and $\alpha_{inv}$ and radar albedos and $\mu_{c}$ were measured for only 14 asteroids. Although the set is small, two sub-groups seen in the $\textit{P}_{min}$ versus $\alpha_{inv}$ plot include asteroids with different radar properties. The asteroids with high radar albedos and small $\mu_{c}$ are located in one sub-group, while the second sub-group contains asteroids with lower radar albedos and higher values of $\mu_{c}$.

  \begin{figure*}[ht!]
        % \vspace*{-1.0 cm}
        \begin{center}
                \includegraphics[width=6.0in]{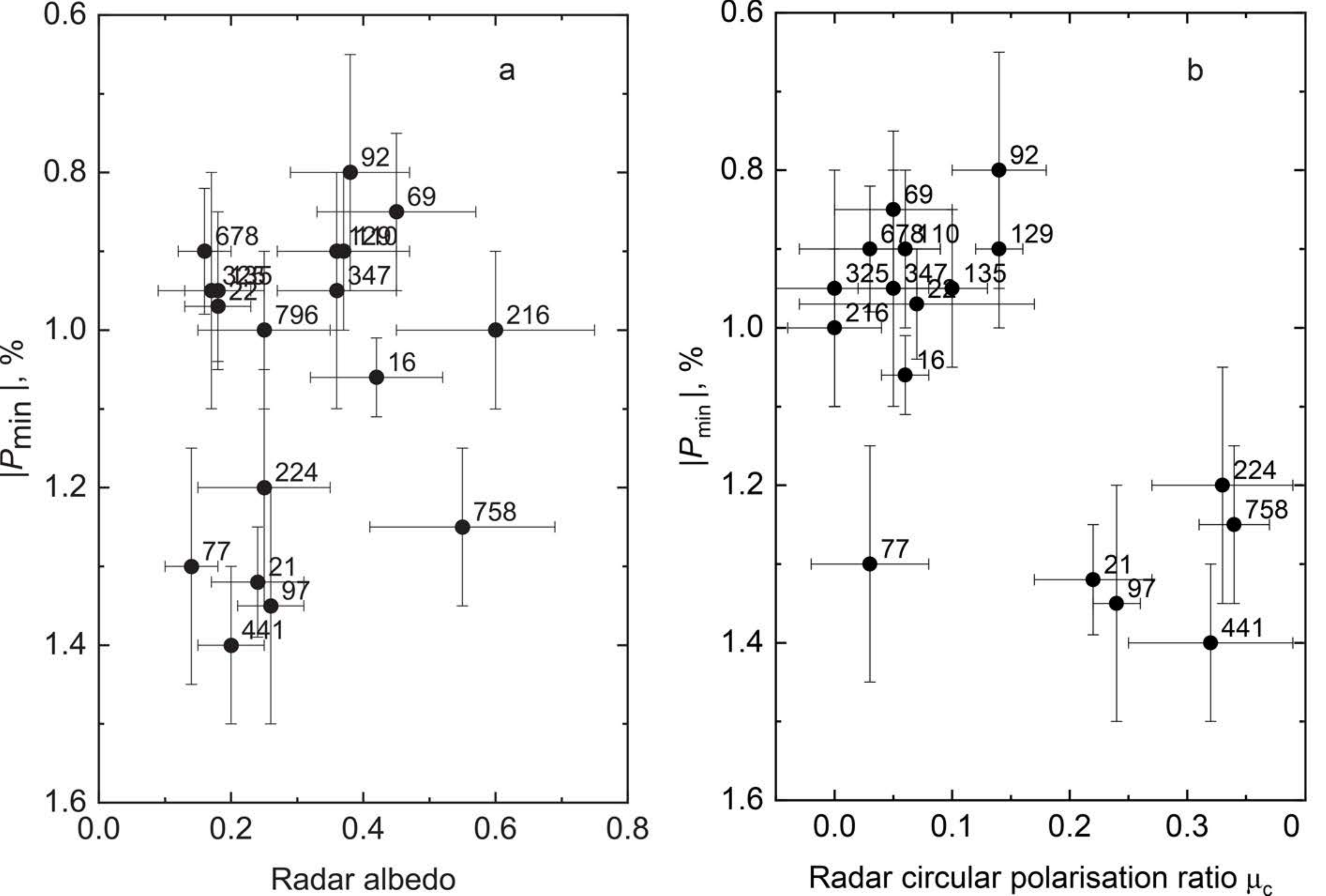} 
                % \vspace*{-1.0 cm}
                \caption{Relationships of $\textit{P}_{min}$ and radar albedos (a) and circular polarisation ratio $\mu_{c}$ (b) for M-type asteroids.}
                \label{fig5}
        \end{center}
\end{figure*}

\subsection{$\textit{P}_{min}$ and spectral slope}

Visible spectra of M-type asteroids are characterised by rather similar behaviour. The difference became noticeable at longer wavelengths, which led to the identification several subclasses among X-complex in Bus-DeMeo taxonomy (\citealt{bus}, \citealt{demeo}). The near-infrared spectra are available for 31 asteroids from our list of 36 asteroids for which $\textit{P}_{min}$ was measured. They were obtained by \citet{clark}, \citet{fornasier}, Ockert-Bell et al. (\citeyear{ockert08}, \citeyear{ockert10}), \citet{hardersen11}, and \citet{neeley}. All these spectra are available at the PDS Asteroid Database\footnote{\url{https://sbn.psi.edu/pds/archive/asteroids.html}}. We used them to calculate spectral slopes in the 1.7-2.4 $\mu$m wavelength range. This range was previously used to characterise near-infrared spectra of M/X type asteroids (e.g. \citealt{fornasier}). We carefully checked each available spectrum for the presence of absorption features. The available spectra are typically featureless in the considered wavelength range. Subtle absorption features reported for several asteroids are present in some spectra but absent from others. We used a simple linear regression to calculate spectral slope and its standard error. In the case of multiple spectra available, we used the most accurate spectra or average value of the slopes. The adopted values of slopes are given in Table 2. They are plotted in Fig.6 versus $|\textit{P}_{min}|$. An inverse correlation is seen between $|\textit{P}_{min}|$ and spectral slope in the 1.7-2.4 $\mu$m wavelength range. Asteroids with shallower spectral slopes at the near-infrared wavelengths have deeper negative polarisation branches. Such correlation implies that polarimetry is rather sensitive to surface composition of asteroids.

 \begin{figure}[ht!]
        \begin{center}
                \includegraphics[width=3.5in]{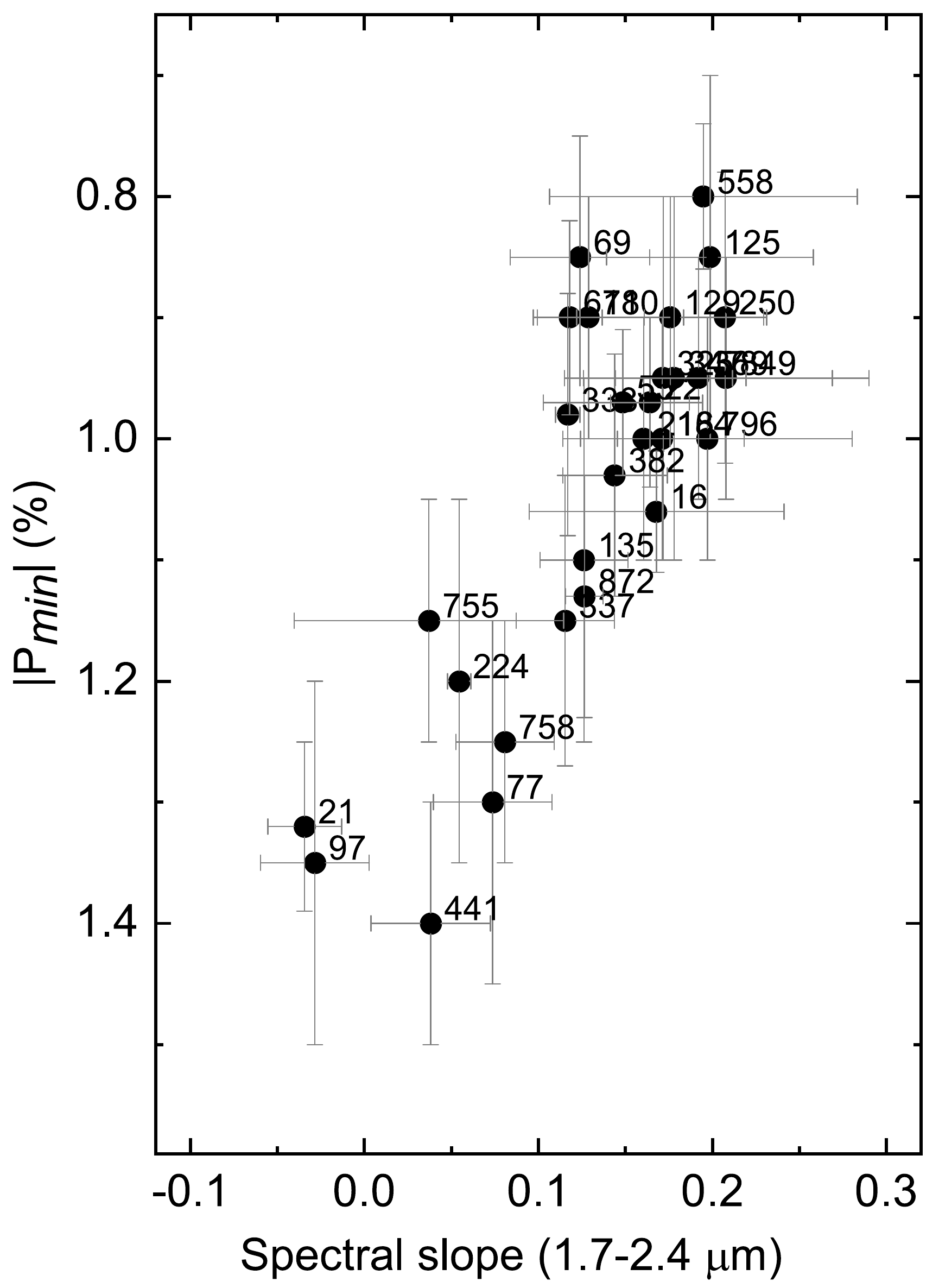} 
                \caption{Relationship of $\textit{P}_{min}$ and spectral slope in the 1.7-2.4 $\mu$m wavelength range (in reflectance/$\mu$m). }
                \label{fig6}
        \end{center}
\end{figure}

\section{Discussion}

New polarimetric observations obtained for 22 asteroids of M/X-type combined with available observations provide a data set of 41 asteroids for which at least one of polarimetric parameters $\textit{P}_{min}$ or $\alpha_{inv}$ were determined. This data set cover the majority of M/X asteroids that were observed in the previous surveys by spectroscopic and radar techniques (\citealt{shepard15}). So, it becomes possible to add polarimetry as a complimentary technique to previous analysis of M-type asteroids. We did not find noticeable correlation with either the presence or absence of a 3 $\mu$m feature or with radar albedos. The most prominent correlation we revealed is the correlation of the depth of negative polarisation and the slope in the near-infrared wavelength range. We also found that asteroids with a deeper negative polarisation branch in optical measurements tend to have higher circular polarisation ratio $\mu_{c}$. Moreover, the two sub-groups seen in the $\textit{P}_{min}$ versus $\alpha_{inv}$ plot are also well distinguished in the plot of $\textit{P}_{min}$ and radar circular polarisation ratios. Below we discuss possible interpretation of the diversity of polarimetric parameters among M-type asteroids.

\textbf{Asteroids with $|\textit{P}_{min}|\sim$ 0.9--1\% and $\alpha_{inv} \sim$ 20--22\textdegree}\ are (16) Psyche, (22) Kalliope, (55) Pandora, (69) Hesperia, (125) Liberatrix, (129) Antigone, (135) Hertha, (216) Kleopatra, (250) Bettina, (347) Pariana, (369) Aeria, (678) Fredegundis, and (796) Sarita. All these asteroids have shown an absorption feature near 0.9 $\mu$m, although its centre and depth derived by different authors did not always match (\citealt{hardersen05}, \citeyear{hardersen11}, \citealt{fornasier}, \citealt{ockert08}, \citeyear{ockert10}). For some of these asteroids, the presence of absorption features in the near-infrared wavelength range was reported, but again the data for the same asteroid by different authors were controversial (see \citealt{hardersen11}). Possible meteorite analogues of asteroids with an absorption feature near 0.9 $\mu$m include mesosiderites (\citealt{vernazza09}, \citealt{hardersen11}) or silicate-bearing iron meteorites (\citealt{hardersen11}). The bulk density estimations of the largest asteroids from this group are comparable to mesosiderites  (\citealt{vernazza21}), but the authors considered stony-iron meteorites as unlikely analogues of these asteroids (\citealt{vernazza21}). All measured main-belt asteroids with high radar albedos (\citealt{shepard15}) also belong to this group.

\textbf{Asteroids with $|\textit{P}_{min}|\geqslant$ 1.2\% and $\alpha_{inv}$ > 22\textdegree}\ are (21) Lutetia, (77) Frigga, (97) Klotho, (441) Bathilde, and most probably (224) Oceana and (758) Mancunia, for which only $\textit{P}_{min}$ was measured. These asteroids have flattened spectra in the near-infrared wavelength, have not revealed an absorption band at 0.9 $\mu$m, have low or moderate radar albedo (except (758) Mancunia) and high circular polarisation ratios (except (77) Frigga). Among these asteroids, (21) Lutetia was well studied during the fly-by of the Rosetta space mission. Some types of carbonaceous chondrites or enstatite chondrites are considered as Lutetia’s surface analogues (\citealt{barucci}). Most probably, all other asteroids in this group also have a similar analogous among meteorites. This conclusion is also supported by laboratory measurements. The measured samples of carbonaceous chondrites and the enstatite chondrite E4 Abee revealed a deeper negative polarisation branch compared to that for iron meteorites (\citealt{zellner77}, \citealt{doll79}, \citealt{lupish89}). Also, enstatite chondrites tend to have shallower spectral continuum slopes than those of iron meteorites (e.g. \citealt{clark}). 

\textbf{Asteroids with extreme polarimetric properties}
include (785) Zwetana and (757) Portlandia. Zwetana is characterised by the deepest negative polarisation branch in our sample. This asteroid revealed several interesting features that placed it outside of typical M types. Available near-infrared spectra demonstrated variable spectral slopes and absorption features detected by some authors (\citealt{clark}, \citealt{ockert10}) but not confirmed by others (\citealt{fornasier}, \citealt{hardersen11}). The considerable variability was also found in radar measurements of (785) Zwetana (\citealt{shepard08}), which was explained by an unusual shape or surface structures of this asteroid (\citealt{shepard15}). The polarimetric observations also revealed an unusually large scatter of data (see Fig.2). Further observations of (785) Zwetana are needed to explain  its unusual surface properties. 
\par Asteroid (757) Portlandia has an extremely small inversion angle previously observed only for low-albedo F-type asteroids (\citealt{belskaya05}). Accurate measurements of its albedo are needed to check whether it belongs to F-type asteroids or if it has unique properties.

Based on a comparison of polarimetric parameters and available information on surface composition of M-type asteroids from other sources, we assume that the sub-groups in the '$\textit{P}_{min}$ -- $\alpha_{inv}$' relationship most likely indicate two different surface compositions. Although it is well known that the value of an inversion angle is sensitive to a regolith particle size increasing as the size of particles decreases (e.g. \citealt{doll89}), such interpretation is ambiguous. The inversion angle also depends on the surface composition increasing with an increase of the refractive index of the surface material (e.g. see \citealt{gilgar}). We know an impressive example of the sensitivity of polarimetric parameters to asteroid surface composition. The discovery of anomalous polarimetric properties of asteroid (234) Barbara (\citealt{cellino06}) led to a separate group of 'barbarians' being established. It was later proved that all asteroids of this group belong to the same L-type (\citealt{devogele}). Asteroids of other taxonomic types are also grouped in the '$\textit{P}_{min}$ -- $\alpha_{inv}$' diagram (\citealt{belskaya17}). Thus, we suggest that two sub-groups of  M-type asteroids with different polarimetric characteristics may indicate their different surface compositions, corresponding to two different meteorite analogues, such as (1) irons and stony-irons, and (2) enstatite and high-iron carbonaceous chondrites. Of course, only with additional data it will be possible to confirm or reject this assumption. 

\section{Conclusions}
We obtained new polarimetric observations for 22 M/X-type asteroids and combined them with previously available observations to determine polarimetric parameters characterising the depth and width of a negative polarisation branch. At least one of these polarimetric parameters were determined for 41 asteroids, which cover the majority of M/X-type asteroids observed in the previous surveys by spectroscopic and radar techniques. The main conclusions from the analysis of polarimetric data and other available data on M-type asteroids can be summarised as follows:
 \begin{enumerate}
      \item M-type asteroids have a wider range of polarimetric parameters as compared to other types, which confirmed a diversity within M-type. Majority of considered asteroids fall into two sub-groups in the plot of polarimetric parameters $\textit{P}_{min}$ and $\alpha_{inv}$. Asteroids (757) Portlandia and (785) Zwetana present extreme cases and most probably do not belong to M-type asteroids.
      \item     We found a correlation of the depth of negative polarisation and the spectral slope in the near-infrared wavelength range. Asteroids with shallower spectral slopes at the near-infrared wavelengths have deeper negative polarisation branches. 
      \item     Asteroids with deeper negative polarisation branches tend to have higher radar circular polarisation ratios $\mu_{c}$. Two sub-groups are also well distinguished in the plot of $\textit{P}_{min}$ and radar circular polarisation ratios.
      \item     We suggest that two sub-groups of  M-type asteroids with different polarimetric characteristics may have different meteorite analogues, such as (1) irons and stony-irons, and (2) enstatite and high-iron carbonaceous chondrites. 
 \end{enumerate}
Thus, polarimetry can be an efficient tool in study of X/M-type asteroids. The accuracy of measurements of polarisation degree should be better than 0.1\% in order to reveal differences in the polarisation phase curve behaviours of individual asteroids. Further observations are needed to increase the statistics and confirmed our findings.

\begin{acknowledgements}
 Ukrainian team is supported by the National Research Foundation of Ukraine, grant N 2020.02/0371 “Metallic asteroids: search for parent bodies of iron meteorites, sources of extraterrestrial resources”. Dipol-2 was built in the cooperation between the University of Turku, Finland, and the Leibniz Institute for Solar Physics, Germany, with the support by the Leibniz Association grant SAW-2011-KIS-7.  We are grateful to the Institute for Astronomy, University of Hawaii for the observing time allocated for us on the T60 telescope. RGH gratefully acknowledges financial support by CONICET through PIP 112-202001-01227 and San Juan National University by a CICITCA grant for the period 2020-2021. TB acknowledges  financial support by the National Science Fund in Bulgaria through contract DN 18/13-12.12.2017. We are grateful to all the defenders of Ukraine from the Russian invasion so that we can safely finalize this article. 
\end{acknowledgements}

\bibliographystyle{aa} 
\bibliography{aapol} 

\begin{thebibliography}{53}
\expandafter\ifx\csname natexlab\endcsname\relax\def\natexlab#1{#1}\fi

\bibitem[{{Al{\'\i}-Lagoa} {et~al.}(2018){Al{\'\i}-Lagoa}, {M{\"u}ller},
  {Usui}, \& {Hasegawa}}]{ali}
{Al{\'\i}-Lagoa}, V., {M{\"u}ller}, T.~G., {Usui}, F., \& {Hasegawa}, S. 2018,
  \aap, 612, A85

\bibitem[{{Bagnulo} {et~al.}(2006){Bagnulo}, {Boehnhardt}, {Muinonen},
  {Kolokolova}, {Belskaya}, \& {Barucci}}]{bagnulo06}
{Bagnulo}, S., {Boehnhardt}, H., {Muinonen}, K., {et~al.} 2006, \aap, 450, 1239

\bibitem[{{Barucci} {et~al.}(2012){Barucci}, {Belskaya}, {Fornasier},
  {Fulchignoni}, {Clark}, {Coradini}, {Capaccioni}, {Dotto}, {Birlan},
  {Leyrat}, {Sierks}, {Thomas}, \& {Vincent}}]{barucci}
{Barucci}, M.~A., {Belskaya}, I.~N., {Fornasier}, S., {et~al.} 2012, \planss,
  66, 23

\bibitem[{{Belskaya} {et~al.}(2017){Belskaya}, {Fornasier}, {Tozzi},
  {Gil-Hutton}, {Cellino}, {Antonyuk}, {Krugly}, {Dovgopol}, \&
  {Faggi}}]{belskaya17}
{Belskaya}, I.~N., {Fornasier}, S., {Tozzi}, G.~P., {et~al.} 2017, \icarus,
  284, 30

\bibitem[{{Belskaya} {et~al.}(2009){Belskaya}, {Levasseur-Regourd}, {Cellino},
  {Efimov}, {Shakhovskoy}, {Hadamcik}, \& {Bendjoya}}]{belskaya}
{Belskaya}, I.~N., {Levasseur-Regourd}, A.-C., {Cellino}, A., {et~al.} 2009,
  \icarus, 199, 97

\bibitem[{{Belskaya} {et~al.}(2005){Belskaya}, {Shkuratov}, {Efimov},
  {Shakhovskoy}, {Gil-Hutton}, {Cellino}, {Zubko}, {Ovcharenko}, {Bondarenko},
  {Shevchenko}, {Fornasier}, \& {Barbieri}}]{belskaya05}
{Belskaya}, I.~N., {Shkuratov}, Y.~G., {Efimov}, Y.~S., {et~al.} 2005, \icarus,
  178, 213

\bibitem[{{Benner} {et~al.}(2008){Benner}, {Ostro}, {Magri}, {Nolan}, {Howell},
  {Giorgini}, {Jurgens}, {Margot}, {Taylor}, {Busch}, \& {Shepard}}]{benner}
{Benner}, L. A.~M., {Ostro}, S.~J., {Magri}, C., {et~al.} 2008, \icarus, 198,
  294

\bibitem[{{Bus} \& {Binzel}(2002)}]{bus}
{Bus}, S.~J. \& {Binzel}, R.~P. 2002, \icarus, 158, 146

\bibitem[{{Cellino} {et~al.}(2016){Cellino}, {Ammannito}, {Magni},
  {Gil-Hutton}, {Tedesco}, {Belskaya}, {De Sanctis}, {Schr{\"o}der},
  {Preusker}, \& {Manara}}]{cellino16}
{Cellino}, A., {Ammannito}, E., {Magni}, G., {et~al.} 2016, \mnras, 456, 248

\bibitem[{{Cellino} {et~al.}(2015){Cellino}, {Bagnulo}, {Gil-Hutton}, {Tanga},
  {Ca{\~n}ada-Assandri}, \& {Tedesco}}]{cellino}
{Cellino}, A., {Bagnulo}, S., {Gil-Hutton}, R., {et~al.} 2015, \mnras, 451,
  3473

\bibitem[{{Cellino} {et~al.}(2006){Cellino}, {Belskaya}, {Bendjoya}, {Di
  Martino}, {Gil-Hutton}, {Muinonen}, \& {Tedesco}}]{cellino06}
{Cellino}, A., {Belskaya}, I.~N., {Bendjoya}, P., {et~al.} 2006, \icarus, 180,
  565

\bibitem[{{Chapman} {et~al.}(1975){Chapman}, {Morrison}, \&
  {Zellner}}]{chapman}
{Chapman}, C.~R., {Morrison}, D., \& {Zellner}, B. 1975, \icarus, 25, 104

\bibitem[{{Clark} {et~al.}(2004){Clark}, {Bus}, {Rivkin}, {Shepard}, \&
  {Shah}}]{clark}
{Clark}, B.~E., {Bus}, S.~J., {Rivkin}, A.~S., {Shepard}, M.~K., \& {Shah}, S.
  2004, \aj, 128, 3070

\bibitem[{{DeMeo} {et~al.}(2009){DeMeo}, {Binzel}, {Slivan}, \& {Bus}}]{demeo}
{DeMeo}, F.~E., {Binzel}, R.~P., {Slivan}, S.~M., \& {Bus}, S.~J. 2009,
  \icarus, 202, 160

\bibitem[{{Devog{\`e}le} {et~al.}(2018){Devog{\`e}le}, {Tanga}, {Cellino},
  {Bendjoya}, {Rivet}, {Surdej}, {Vernet}, {Sunshine}, {Bus}, {Abe}, {Bagnulo},
  {Borisov}, {Campins}, {Carry}, {Licandro}, {McLean}, \&
  {Pinilla-Alonso}}]{devogele}
{Devog{\`e}le}, M., {Tanga}, P., {Cellino}, A., {et~al.} 2018, \icarus, 304, 31

\bibitem[{{Dollfus} {et~al.}(1979){Dollfus}, {Mandeville}, \&
  {Duseaux}}]{doll79}
{Dollfus}, A., {Mandeville}, J.~C., \& {Duseaux}, M. 1979, \icarus, 37, 124

\bibitem[{{Dollfus} {et~al.}(1989){Dollfus}, {Wolff}, {Geake}, {Lupishko}, \&
  {Dougherty}}]{doll89}
{Dollfus}, A., {Wolff}, M., {Geake}, J.~E., {Lupishko}, D.~F., \& {Dougherty},
  L.~M. 1989, in Asteroids II, ed. R.~P. {Binzel}, T.~{Gehrels}, \& M.~S.
  {Matthews}, 594--616

\bibitem[{{Elkins-Tanton} {et~al.}(2020){Elkins-Tanton}, {Asphaug}, {Bell},
  {Bercovici}, {Bills}, {Binzel}, {Bottke}, {Dibb}, {Lawrence}, {Marchi},
  {McCoy}, {Oran}, {Park}, {Peplowski}, {Polanskey}, {Prettyman}, {Russell},
  {Schaefer}, {Weiss}, {Wieczorek}, {Williams}, \& {Zuber}}]{elkins}
{Elkins-Tanton}, L.~T., {Asphaug}, E., {Bell}, J.~F., {et~al.} 2020, Journal of
  Geophysical Research (Planets), 125, e06296

\bibitem[{{Fornasier} {et~al.}(2011){Fornasier}, {Clark}, \&
  {Dotto}}]{fornasier11}
{Fornasier}, S., {Clark}, B.~E., \& {Dotto}, E. 2011, \icarus, 214, 131

\bibitem[{{Fornasier} {et~al.}(2010){Fornasier}, {Clark}, {Dotto},
  {Migliorini}, {Ockert-Bell}, \& {Barucci}}]{fornasier}
{Fornasier}, S., {Clark}, B.~E., {Dotto}, E., {et~al.} 2010, \icarus, 210, 655

\bibitem[{{Gil-Hutton}(2007)}]{gilhutton}
{Gil-Hutton}, R. 2007, \aap, 464, 1127

\bibitem[{{Gil-Hutton}(2017)}]{gil}
{Gil-Hutton}, R. 2017, {Catalogue of asteroid polarization curves, presented at
  "Asteroid, Comets, Meteors 2017", Montevideo, Uruguay}

\bibitem[{{Gil-Hutton} \& {Garc{\'\i}a-Migani}(2017)}]{gilgar}
{Gil-Hutton}, R. \& {Garc{\'\i}a-Migani}, E. 2017, \aap, 607, A103

\bibitem[{{Gil-Hutton} {et~al.}(2017){Gil-Hutton}, {L{\'o}pez-Sisterna}, \&
  {Calandra}}]{gil17}
{Gil-Hutton}, R., {L{\'o}pez-Sisterna}, C., \& {Calandra}, M.~F. 2017, \aap,
  599, A114

\bibitem[{{Hardersen} {et~al.}(2011){Hardersen}, {Cloutis}, {Reddy},
  {Moth{\'e}-Diniz}, \& {Emery}}]{hardersen11}
{Hardersen}, P.~S., {Cloutis}, E.~A., {Reddy}, V., {Moth{\'e}-Diniz}, T., \&
  {Emery}, J.~P. 2011, Meteorit Planet Sci, 46, 1910

\bibitem[{{Hardersen} {et~al.}(2005){Hardersen}, {Gaffey}, \&
  {Abell}}]{hardersen05}
{Hardersen}, P.~S., {Gaffey}, M.~J., \& {Abell}, P.~A. 2005, \icarus, 175, 141

\bibitem[{{Jockers} {et~al.}(2000){Jockers}, {Credner}, {Bonev}, {Kisele},
  {Korsun}, {Kulyk}, {Rosenbush}, {Andrienko}, {Karpov}, {Sergeev}, \&
  {Tarady}}]{jockers}
{Jockers}, K., {Credner}, T., {Bonev}, T., {et~al.} 2000, Kinematika i Fizika
  Nebesnykh Tel Supplement, 3, 13

\bibitem[{{Kaasalainen} {et~al.}(2003){Kaasalainen}, {Piironen}, {Kaasalainen},
  {Harris}, {Muinonen}, \& {Cellino}}]{kaasa}
{Kaasalainen}, S., {Piironen}, J., {Kaasalainen}, M., {et~al.} 2003, \icarus,
  161, 34

\bibitem[{{Lazzaro} {et~al.}(2004){Lazzaro}, {Angeli}, {Carvano},
  {Moth{\'e}-Diniz}, {Duffard}, \& {Florczak}}]{lazzaro}
{Lazzaro}, D., {Angeli}, C.~A., {Carvano}, J.~M., {et~al.} 2004, \icarus, 172,
  179

\bibitem[{{Lupishko}(2019)}]{lupishko}
{Lupishko}, D. 2019, NASA Planetary Data System, 1

\bibitem[{{Lupishko}(2018)}]{lupishko18}
{Lupishko}, D.~F. 2018, Solar System Research, 52, 98

\bibitem[{{Lupishko} \& {Belskaya}(1989)}]{lupish89}
{Lupishko}, D.~F. \& {Belskaya}, I.~N. 1989, \icarus, 78, 395

\bibitem[{{Magalhaes} {et~al.}(1996){Magalhaes}, {Rodrigues}, {Margoniner},
  {Pereyra}, \& {Heathcote}}]{magalhaes}
{Magalhaes}, A.~M., {Rodrigues}, C.~V., {Margoniner}, V.~E., {Pereyra}, A., \&
  {Heathcote}, S. 1996, in Astronomical Society of the Pacific Conference
  Series, Vol.~97, Polarimetry of the Interstellar Medium, ed. W.~G. {Roberge}
  \& D.~C.~B. {Whittet}, 118

\bibitem[{{Mainzer} {et~al.}(2016){Mainzer}, {Bauer}, {Cutri}, {Grav},
  {Kramer}, {Masiero}, {Nugent}, {Sonnett}, {Stevenson}, \& {Wright}}]{mainzer}
{Mainzer}, A.~K., {Bauer}, J.~M., {Cutri}, R.~M., {et~al.} 2016, NASA Planetary
  Data System, EAR

\bibitem[{{Masiero} {et~al.}(2011){Masiero}, {Mainzer}, {Grav}, {Bauer},
  {Cutri}, {Dailey}, {Eisenhardt}, {McMillan}, {Spahr}, {Skrutskie}, {Tholen},
  {Walker}, {Wright}, {DeBaun}, {Elsbury}, {Gautier}, {Gomillion}, \&
  {Wilkins}}]{masiero}
{Masiero}, J.~R., {Mainzer}, A.~K., {Grav}, T., {et~al.} 2011, \apj, 741, 68

\bibitem[{{Muinonen} {et~al.}(2009){Muinonen}, {Penttil{\"a}}, {Cellino},
  {Belskaya}, {Delb{\`o}}, {Levasseur-Regourd}, \& {Tedesco}}]{muinonen}
{Muinonen}, K., {Penttil{\"a}}, A., {Cellino}, A., {et~al.} 2009, Meteorit
  Planet Sci, 44, 1937

\bibitem[{{Neeley} {et~al.}(2014){Neeley}, {Clark}, {Ockert-Bell}, {Shepard},
  {Conklin}, {Cloutis}, {Fornasier}, \& {Bus}}]{neeley}
{Neeley}, J.~R., {Clark}, B.~E., {Ockert-Bell}, M.~E., {et~al.} 2014, \icarus,
  238, 37

\bibitem[{{Ockert-Bell} {et~al.}(2010){Ockert-Bell}, {Clark}, {Shepard},
  {Isaacs}, {Cloutis}, {Fornasier}, \& {Bus}}]{ockert10}
{Ockert-Bell}, M.~E., {Clark}, B.~E., {Shepard}, M.~K., {et~al.} 2010, \icarus,
  210, 674

\bibitem[{{Ockert-Bell} {et~al.}(2008){Ockert-Bell}, {Clark}, {Shepard},
  {Rivkin}, {Binzel}, {Thomas}, {DeMeo}, {Bus}, \& {Shah}}]{ockert08}
{Ockert-Bell}, M.~E., {Clark}, B.~E., {Shepard}, M.~K., {et~al.} 2008, \icarus,
  195, 206

\bibitem[{{Ostro} {et~al.}(1985){Ostro}, {Campbell}, \& {Shapiro}}]{ostro}
{Ostro}, S.~J., {Campbell}, D.~B., \& {Shapiro}, I.~I. 1985, Science, 229, 442

\bibitem[{{Piirola} {et~al.}(2014){Piirola}, {Berdyugin}, \&
  {Berdyugina}}]{piirola}
{Piirola}, V., {Berdyugin}, A., \& {Berdyugina}, S. 2014, in Society of
  Photo-Optical Instrumentation Engineers (SPIE) Conference Series, Vol. 9147,
  Ground-based and Airborne Instrumentation for Astronomy V, ed. S.~K.
  {Ramsay}, I.~S. {McLean}, \& H.~{Takami}, 91478I

\bibitem[{{Piirola} {et~al.}(2020){Piirola}, {Berdyugin}, {Frisch}, {Kagitani},
  {Sakanoi}, {Berdyugina}, {Cole}, {Harlingten}, \& {Hill}}]{piirola20}
{Piirola}, V., {Berdyugin}, A., {Frisch}, P.~C., {et~al.} 2020, \aap, 635, A46

\bibitem[{{Rivkin} {et~al.}(1995){Rivkin}, {Howell}, {Britt}, {Lebofsky},
  {Nolan}, \& {Branston}}]{rivkin}
{Rivkin}, A.~S., {Howell}, E.~S., {Britt}, D.~T., {et~al.} 1995, \icarus, 117,
  90

\bibitem[{{Rivkin} {et~al.}(2000){Rivkin}, {Howell}, {Lebofsky}, {Clark}, \&
  {Britt}}]{rivkin00}
{Rivkin}, A.~S., {Howell}, E.~S., {Lebofsky}, L.~A., {Clark}, B.~E., \&
  {Britt}, D.~T. 2000, \icarus, 145, 351

\bibitem[{{Shepard} {et~al.}(2008){Shepard}, {Clark}, {Nolan}, {Howell},
  {Magri}, {Giorgini}, {Benner}, {Ostro}, {Harris}, {Warner}, {Pray}, {Pravec},
  {Fauerbach}, {Bennett}, {Klotz}, {Behrend}, {Correia}, {Coloma}, {Casulli},
  \& {Rivkin}}]{shepard08}
{Shepard}, M.~K., {Clark}, B.~E., {Nolan}, M.~C., {et~al.} 2008, \icarus, 195,
  184

\bibitem[{{Shepard} {et~al.}(2010){Shepard}, {Clark}, {Ockert-Bell}, {Nolan},
  {Howell}, {Magri}, {Giorgini}, {Benner}, {Ostro}, {Harris}, {Warner},
  {Stephens}, \& {Mueller}}]{shepard10}
{Shepard}, M.~K., {Clark}, B.~E., {Ockert-Bell}, M., {et~al.} 2010, \icarus,
  208, 221

\bibitem[{{Shepard} {et~al.}(2015){Shepard}, {Taylor}, {Nolan}, {Howell},
  {Springmann}, {Giorgini}, {Warner}, {Harris}, {Stephens}, {Merline},
  {Rivkin}, {Benner}, {Coley}, {Clark}, {Ockert-Bell}, \& {Magri}}]{shepard15}
{Shepard}, M.~K., {Taylor}, P.~A., {Nolan}, M.~C., {et~al.} 2015, \icarus, 245,
  38

\bibitem[{{Tholen}(1989)}]{tholen}
{Tholen}, D.~J. 1989, in Asteroids II, ed. R.~P. {Binzel}, T.~{Gehrels}, \&
  M.~S. {Matthews}, 1139--1150

\bibitem[{{Usui} {et~al.}(2011){Usui}, {Kuroda}, {M{\"u}ller}, {Hasegawa},
  {Ishiguro}, {Ootsubo}, {Ishihara}, {Kataza}, {Takita}, {Oyabu}, {Ueno},
  {Matsuhara}, \& {Onaka}}]{usui}
{Usui}, F., {Kuroda}, D., {M{\"u}ller}, T.~G., {et~al.} 2011, \pasj, 63, 1117

\bibitem[{{Vernazza} {et~al.}(2009){Vernazza}, {Brunetto}, {Binzel}, {Perron},
  {Fulvio}, {Strazzulla}, \& {Fulchignoni}}]{vernazza09}
{Vernazza}, P., {Brunetto}, R., {Binzel}, R.~P., {et~al.} 2009, \icarus, 202,
  477

\bibitem[{{Vernazza} {et~al.}(2021){Vernazza}, {Ferrais}, {Jorda},
  {Hanu{\v{s}}}, {Carry}, {Marsset}, {Bro{\v{z}}}, {Fetick}, {Viikinkoski},
  {Marchis}, {Vachier}, {Drouard}, {Fusco}, {Birlan}, {Podlewska-Gaca},
  {Rambaux}, {Neveu}, {Bartczak}, {Dudzi{\'n}ski}, {Jehin}, {Beck}, {Berthier},
  {Castillo-Rogez}, {Cipriani}, {Colas}, {Dumas}, {{\v{D}}urech}, {Grice},
  {Kaasalainen}, {Kryszczynska}, {Lamy}, {Le Coroller}, {Marciniak},
  {Michalowski}, {Michel}, {Santana-Ros}, {Tanga}, {Vigan}, {Witasse}, {Yang},
  {Antonini}, {Audejean}, {Aurard}, {Behrend}, {Benkhaldoun}, {Bosch},
  {Chapman}, {Dalmon}, {Fauvaud}, {Hamanowa}, {Hamanowa}, {His}, {Jones},
  {Kim}, {Kim}, {Krajewski}, {Labrevoir}, {Leroy}, {Livet}, {Molina},
  {Montaigut}, {Oey}, {Payre}, {Reddy}, {Sabin}, {Sanchez}, \&
  {Socha}}]{vernazza21}
{Vernazza}, P., {Ferrais}, M., {Jorda}, L., {et~al.} 2021, \aap, 654, A56

\bibitem[{{Zellner} \& {Gradie}(1976)}]{zellner}
{Zellner}, B. \& {Gradie}, J. 1976, \aj, 81, 262

\bibitem[{{Zellner} {et~al.}(1977){Zellner}, {Leake}, {Lebertre}, {Duseaux}, \&
  {Dollfus}}]{zellner77}
{Zellner}, B., {Leake}, M., {Lebertre}, T., {Duseaux}, M., \& {Dollfus}, A.
  1977, Lunar and Planetary Science Conference Proceedings, 1, 1091

\end{thebibliography}

\end{document}